# Geographical Space Based on Urban Allometry and Fractal Dimension

Yanguang Chen

(Department of Geography, College of Urban and Environmental Sciences, Peking University, Beijing 100871, P.R. China. E-mail: chenyg@pku.edu.cn)

**Abstract**: The conventional concept of geographical space is mainly referred to actual space based on landscape, maps, and remote sensing images. However, this notion of space is not enough to interpret different types of fractal dimension of cities. The fractal dimensions derived from Zipf's law and time series analysis do not belong to the traditional geographical space. Based on the nature of the datasets, the urban allometry can be divided into three types: longitudinal allometry indicating time, transversal allometry indicating hierarchy, and isoline allometry indicating space. According to the principle of dimension consistency, an allometric scaling exponent must be a ratio of one fractal dimension to another. From abovementioned three allometric models, we can derive three sets of fractal dimension. In light of the three sets of fractal dimension and the principle of dimension uniqueness, urban geographical space falls into three categories, including the real space based on isoline allometry and spatial distribution, the phase space based on longitudinal allometry and time series, and order space based on transversal allometry and rank-size distribution. The generalized space not only helps to explain various fractal dimensions of cities, but also can be used to develop new theory and methods of geospatial analysis.

## 1. Introduction

Three basic laws are important in geographical analysis, that is, distance decay law, rank-size law, and allometric growth law. The first law is a spatial law, the second law is a hierarchical law, and third law is originally a temporal law. Nowadays, the law of allometric growth has been generalized to allometric scaling law and can be used to associate space and time with hierarchy. In urban studies, allometric scaling law has been employed to describe urban growth, urban form, and urban systems. The law of allometric growth initially implies that the rate of relative growth of an organ is a constant fraction of the rate of relative growth of the total organism (Beckmann, 1958; Lee, 1989). In general system theory, allometry means a constant ratio of one relative growth rate to another relative growth rate (Bertalanffy, 1968). We can find two types of allometric scaling relations: First, the ratio of the relative growth rate of a part to the relative growth rate of the whole is a constant. For example, the relationships between a central city and a system of cities (Beckmann, 1958; Chen, 2017). Second, the ratio of the relative growth rate of one part to the relative growth rate of another part is a constant. For example, the relationships between urban perimeter and urban area (Batty and Longley, 1994; Benguigui *et al*, 2006), the relationships between urban area and population (Batty and Longley, 1994; Lee, 1989), the relationships between two cities (Chen, 2017), and so on (West, 2017).

The question is how to interpret the allometric scaling exponent. This involves two basic principles about dimension, that is, dimension consistency principle and dimension uniqueness principle. Allometric relation always takes on power laws, and a power law is in essence a geometric measure relation, which obeys the principle of dimensional consistency. In this sense, an allometric scaling exponent is a ratio of one dimension value to another dimension value. Unfortunately, it is hard to explain the empirical values of allometric scaling exponents using the concepts from traditional mathematics. It is fractal geometry rather than Euclidean geometry that can be adopted to effectively interpret the allometric scaling exponents in scientific research (Batty and Longley, 1994; Chen and Xu, 1999; Chen, 2010; West *et al*, 1997; West *et al*, 1999). On the other hand, dimension is a kind of geometric characteristic quantity of space. A dimension value corresponds to a spatial form. A geometric object and an aspect of the object bear only one dimension value. This is the principle of dimension uniqueness. However, in geographical analysis, we can obtain several dimension values for the same geographical object. To solve the problem of dimension paradox, we necessarily reconsider the concepts of geographical space. This work is devoted to deriving three types of geographical space from different allometric scaling relations. With the help of new results of space classification, some specious problems in geographical research can be clarified, including the confusion between the box dimension and the similarity dimension of an urban system.

## 2. Allometry models of cities

### 2.1 Allometric scaling classification

Allometric scaling relations can be divided into three categories, that is, longitudinal allometry, transversal allometry, and isoline allometry. The transversal allometry can be equivalently expressed by cross-sectional allometry and hierarchical allometry (Table 1). The urban area-population allometric growth is a simple and good example to illustrate the three types of geographical allometric relations. Using the allometric scaling relation between urban area and population size, we can derive three concepts of geographical space. The urban area-population allometric relation is well known for geographers (Nordbeck, 1971; Lo and Welch, 1977). The model can be formulated as

$$A = aP^b = aP^{D_a/D_p} \,, \tag{1}$$

in which $a$ refers to the proportionality coefficient, and $b=D_a/D_p$ to the allometric scaling exponent (Lee, 1989). According to the principle of dimension consistency, the allometric exponent is a ratio of two fractal dimensions (Batty and Longley, 1994; Chen, 2010), that is

$$b = \frac{D_a}{D_p} \,, \tag{2}$$

where $D_a$ refers to the fractal dimension of urban area, and $D_p$ to the fractal dimension of urban population. The allometric scaling relation between urban area and population size can be examined from three angles of view.

The first is the isoline allometry based on spatial distribution, which reflects the spatial pattern of urban form. For a given city at a certain time, equations (1) and (2) should be replaced by

$$A(r) = a'P(r)^{b'} = a'P(r)^{D_a'/D_p'} \,, \tag{3}$$

where $r$ denotes the radius from the city center, $A(r)$ refers to the land-use area within a radius of $r$ unit from the center ($0 \leq r \leq R$, where $R$ is the maximum radius of a cities), and $P(r)$ to the population within the same sphere as $A(r)$, $a'$ is the proportionality coefficient, and $b'=D_a'/D_p'$ is the scaling exponent. Equation (3) can be derived from two fractal models as follows

$$A(r) = A_1 r^{D_a'} \,, \quad P(r) = P_1 r^{D_p'} \,, \tag{4}$$

where $A_1$ and $P_1$ are two proportionality constants, $D_a'$ is the fractal dimension of urban land use form, and $D_p'$ is the fractal dimension of population distribution of the city. This suggests that the fractal dimensions $D_a'$ and $D_p'$ belong to the real geographical space ($0 \leq D_a', D_p' \leq 2$), which is always confined in a 2-dimensional Euclidean space by maps or digital maps (Chen *et al*, 2019).

The second is the longitudinal allometry based on time series, which reflects the dynamic process of urban growth. For a given city at a certain time, equations (1) and (2) should be replaced by

$$A(t) = a'' P(t)^{b''} = a'' P(t)^{D_a''/D_p''},$$ (5)

where $t$ denotes the time ($t=1, 2, \ldots, n$), $A(t)$ refers to the land-use area in the $t$th time within a radius of $R$ unit from the center, and $P(t)$ to the population in the same time within the same sphere as $A(t)$, $a''$ is the proportionality constant, and $b''=D_a''/D_p''$ is the scaling exponent (Chen, 2010). Equation (5) can be derived from two models such as

$$A(t) = A_T R(t)^{D_a''}, \quad P(t) = P_T R(t)^{D_p''},$$ (6)

where $A_T$ and $P_T$ are two proportionality constants, $R(t)$ is the largest radius of a city in the $t$th year, $D''_a$ is the average fractal dimension of urban land use form in the $n$ year, and $D''_p$ is the average dimension of population distribution of the city in the same period. If the area within an urban envelope is $A$, the largest radius can be defined by $R=F/2=(A/\pi)^{1/2}$, where $F$ denotes Feret's diameter (Batty and Longley, 1994). This implies that the fractal dimensions $D''_a$ and $D''_p$ belong to a generalized geographical space—phase space ($0 \leq D''_a, D''_p \leq 3$), which is always determined with one or more time series.

The third is the transversal allometry based on cross-sectional data, which reflects the rank-size distribution or hierarchical structure of urban systems. For $N$ cities within a region in a given time, equation (1) should be replaced by

$$A(k) = a''' P(k)^{b'''} = a''' P(r)^{D_a'''/D_p'''},$$ (7)

where $k$ denotes the rank of a city ($k=1, 2, \ldots, N$), $A(k)$ refers to the land-use area within an urban envelope, and $P(k)$ to the population size inside the same urban boundary, $a'''$ is the proportionality constant, and $b'''=D_a'''/D_p'''$ is the scaling exponent. Equation (7) can be derived from two power laws

$$A(k) = A_1 R(k)^{D_a'''}, P(k) = P_1 R(k)^{D_p'''},$$ (8)

where $A_1$ and $P_1$ are two proportionality constants, $D_a'''$ is the average fractal dimension of urban land use form of the $N$ cities, and $D_p'''$ is the average dimension of population distribution of the same urban system. This implies that the fractal dimensions $D_a'''$ and $D_p'''$ belong to another generalized geographical space—order space ($0 \leq D_a''', D_p''' \leq 3$), which is always determined by one or more rank-size series. In fact, equation (7) can be equivalently expressed as the following hierarchical scaling relation

$$A(m) = a''' P(m)^{b'''} = a''' P(m)^{D_a'''/D_p'''},\qquad(9)$$

which can be derived from

$$A(m) = A_1 R(m)^{D_a'''},\quad P(m) = P_1 R(m)^{D_p'''},\qquad(10)$$

where $m$=1,2,3,… represents the order number of levels in a hierarchy (Chen, 2010; Chen and Feng, 2017). This suggests that the rank-size series that is/are used to define an order space can be substituted with one or more hierarchical series. What is more, equation (7) can be derived from two Zipf laws, area-based Zipf's law and population-based Zipf's law (Table 1).

## 2.2 Allometric scaling exponents and fractal dimension

According to the principle of dimension consistency, the allometric scaling exponent is a ratio of one fractal dimension to another fractal dimension. In geometry, a measure (e.g., length) is proportional to another measure (e.g., area) if and only if the two measures bear the same dimension. For example, length is not proportional to area, but length is proportional to square root of area. Generally speaking, we have geometric measure relation as below

$$L^{1/1} \propto A^{1/2} \propto V^{1/3},\qquad(11)$$

where $L$, $A$, and $V$ represent length, area, and volume, respectively. Equation (11) can be generalized to fractal measure relation such as (Mandelbrot, 1982; Takayasu, 1990)

$$L^{1/1} \propto A^{1/2} \propto V^{1/3} \propto M^{1/D},\qquad(12)$$

where $M$ refers to generalized volume, and $D$ to fractal dimension. Here Euclidean dimension is regarded as the special case of fractal dimension. Suppose that the dimension of urban area is $d$=2, and the dimension of urban population is $d$=3. According to equation (11), we have a geometric measure relation between one measure $A$ and another measure $V$ such as $A = aV^{2/3}$, where $a$ denotes a proportionality coefficient. According to equation (12), we have a general proportional relation between one measure $A$ and another measure $M$ such as $A = aM^{2/D}$.

In short, an allometric scaling model can be decomposed into two growth processes, or two spatial distributions, or two probability distributions. Based on spatial data, an allometric scaling process can be decomposed into a pair of power law distributions; Based on time series, an allometric growth can be decomposed into a pair of exponential growths or logistical growth; Based on cross-sectional data, an allometric scaling process can be decomposed into a pair of exponential distributions or power law distributions of probability. Thus we have isoline allometry indicative of spatial patterns, longitudinal allometry indicative of temporal process, and transversal allometry indicative of hierarchical structure (Table 1). From each allometry, we can derive a pair of fractal parameters indicating dimension of some types of geographical space.

**Table 1 The spatial, longitudinal, and transversal allometric scaling relations of cities and the related growth or distribution functions**

| Item | Type | Sub-type | Basic models | Main model | Parameters |
|------|------|----------|--------------|-----------|------------|
| Space | Isoline allometry | Spatial allometry | $S(r) = S_1 r^{D_s}$<br>$A(r) = A_1 r^{D_a}$ | $A(r) = aS(r)^b$ | $a = A_1 S_1^{-b}$<br>$b = D_a / D_s$ |
| Time | Longitudinal allometry | Exponential allometry | $S_t = S_0 e^{ut}$<br>$A_t = A_0 e^{vt}$ | $A_t = aS_t^b$ | $a = A_0 S_0^{-b}$<br>$b = v / u$ |
| | | Logistic allometry | $S_t = \dfrac{S_{max}}{1 + (S_{max} / S_0 - 1)e^{-vt}}$<br>$A_t = \dfrac{A_{max}}{1 + (A_{max} / A_0 - 1)e^{-ut}}$ | $\dfrac{A_t}{A_{max} - A_t}$<br>$= a(\dfrac{S_t}{S_{max} - S_t})^b$ | $a = \dfrac{A_0}{A_{max} - A_0}$<br>$\div (\dfrac{S_0}{S_{max} - S_0})^b$<br>$b = v / u$ |
| Hierarchy | Crosssectional allometry | Power allometry | $S_k = S_1 k^{-q}$<br>$A_k = A_1 k^{-p}$ | $A_k = aS_k^b$ | $a = A_1 S_1^{-b}$<br>$b = p / q$ |
| | Hierarchical allometry | Exponential allometry | $S_m = S_1 r_s^{1-m}$<br>$A_m = A_1 r_a^{1-m}$ | $A_m = aS_m^b$ | $a = A_1 S_1^{-b}$<br>$b = \ln r_a / \ln r_s$ |
| | | Power allometry | $S_m = S_1 N_m^{-q}$<br>$A_m = A_1 N_m^{-p}$ | $A_m = aS_m^b$ | $a = A_1 S_1^{-b}$<br>$b = p / q$ |

**Note**: The symbols are as follows: $t$—time; $r$—distance; $k$—rank; $m$—level; $S$—(population) size; $A$—urban area; $a$, $b$, $p$, $q$, $u$, $v$, $r_a$, $r_p$, $A_0$, $A_1$, $A_{max}$, $D_a$, $D_s$, $S_0$, $S_1$, $S_{max}$ are all parameters (proportionality coefficient, fractal dimension, scaling exponent, ratio, capacity, etc.).

## 2.3 Three types of geographical space

Since there are three types of fractal dimensions for a city as a system or a system of cities, the notion of generalized space should be introduced into geography. Given different spatio-temporal conditions, allometric scaling model will suggests three types of urban geographical space: real space, phase space, and order space (Chen, 2014). The first geographical space is the *real space* (R-space). This is the conventional, concrete geographical space, which can be described by field investigation, map, remote sensing image data, etc. Using spatial data of a city or an urban system, we can model it through real space. The second geographical space is the *phase space* (P-space). This is the first abstract geographical space, which can be depicted by one or more time series. Using the observational data of temporal process of a city or an urban system, we can model it through phase space. The third geographical space is the *order space* (O-space). This is the second abstract geographical space, which can be characterized by rank-size series or hierarchical series. Using the cross-sectional data of a city or an urban system, we will be able to model it through order space. Different types of geographical space correspond to different types of fractal dimension (Table 2).

**Table 2 Three types of geographical space: real space, phase space, and order space**

| Space | Description | Physical base and data | Basic fractal dimension | Dimension value range |
|---|---|---|---|---|
| Real space (R-space: the first space) | Empirical space | Spatial series or random observational data based on maps, digital maps, remotely sensed images, etc. | Box dimension, radial dimension, correlation dimension | $0 \leq D \leq 2$ |
| Phase space (P-space: the second space) | Abstract space | Temporal series based on daily/monthly/yearly observations and measurements, etc. | Similarity dimension, correlation dimension | $0 \leq D \leq 3$ |
| Order space (O-space: the third space) | Abstract space | Cross-sectional data based on regional observations and measurements, etc. | Similarity dimension, correlation dimension | $0 \leq D \leq 3$ |

In theory, the same kind of fractal dimension of different spaces should be equal to one another. For a given city at a given time ($t$ is determined), if the urban radius is defined according to certain criterion ($r=R$), we will have

$$b = \frac{D_a'}{D_p'} = \frac{D_a''}{D_p''} = \frac{D_a'''}{D_p'''}. \tag{13}$$

However, because of random disturbance and varied human factors, the observational data do not always support this equation. In practice, equation (13) should be replaced by an approximate relation in the following form

$$b = \frac{D_a'}{D_p'} \approx \frac{D_a''}{D_p''} \approx \frac{D_a'''}{D_p'''}, \tag{14}$$

which can be validated with the statistical average of large-sized samples.

## 2.4 Fractal methods of spatial analysis

In the past, we used distance to characterize geospatial space. If a geographical phenomenon bears characteristic scales, distance will be an effect measure for spatial description of geographical systems. On the contrary, if a geographical phenomenon possesses no characteristic scale, the distance measurement will be invalid for geographical spatial analysis. In this case, the characteristic scale should be replaced by scaling, and distance-based space should be replaced by dimension-based space. Anyway, dimension is the characteristic parameter of space. Euclidean dimension has no more geographical information, but fractal dimension give us useful geographical information for spatial analysis. In short, fractal geometry provide a powerful tool for scaling analysis in geography, especially, in urban studies. We have various fractal parameters, which can be applied to geographical analysis of different types of space (Table 3).

**Table 3 Methods of urban fractal dimension estimation of three types of space based on time series, spatial structure and hierarchical structure**

| Object | Method | Object | Fractal dimension |
|---|---|---|---|
| **R-space for pattern: Spatial structure, texture, and distribution** | Box counting method | Form/network | Box dimension |
| | Sandbox | Growth | Sandbox dimension |
| | Radius-area/number scaling (cluster growing) | Growth | Radial dimension |
| | Wave spectral analysis | Form/network | Image dimension |
| | Walking-divider method | Boundary | Boundary dimension |
| | Perimeter-area scaling | Boundary | Boundary dimension |
| | …… | …… | …… |
| | Power spectral analysis | Process | Self-affine dimension |
| | Reconstructing phase space | Dynamics | Correlation dimension |

| **P-space for process: Time series** | Elasticity relation | Growth | Similarity dimension |
| | …… | …… | …… |
| **O-space for cascade: Hierarchical structure** | Size distribution function | Hierarchy | Similarity dimension |
| | Hierarchical scaling | Hierarchy | Similarity dimension |
| | Allometric scaling | Relation | Dimension ratio |
| | Renormalization | Structure | Network dimension |
| | …… | …… | …… |

# 3. Empirical analysis

## 3.1 Case of R-space

Three typical examples can be presented to illustrate the allometric scaling defined in different types of geographical space. The first case is the spatial allometric relation between urban area and total length of streets based on spatial distribution data. This type of allometry reflects the real space (R-space) of urban geographical systems. Two Chinese cities, Changchun and Jinan, are taken into account (Chen *et al*, 2019). Based on different searching radius $r$, a number of urban envelopes of a city can be identified. Each urban envelope gives an urban area, $A(r)$, and a total length of streets, $L(r)$, of the city. The relationships between urban area and corresponding street length follow spatial allometric scaling law (Figure 1). The scaling exponent is the ratio of the fractal dimension of street network and the fractal dimension of urban form in the real geographical space.

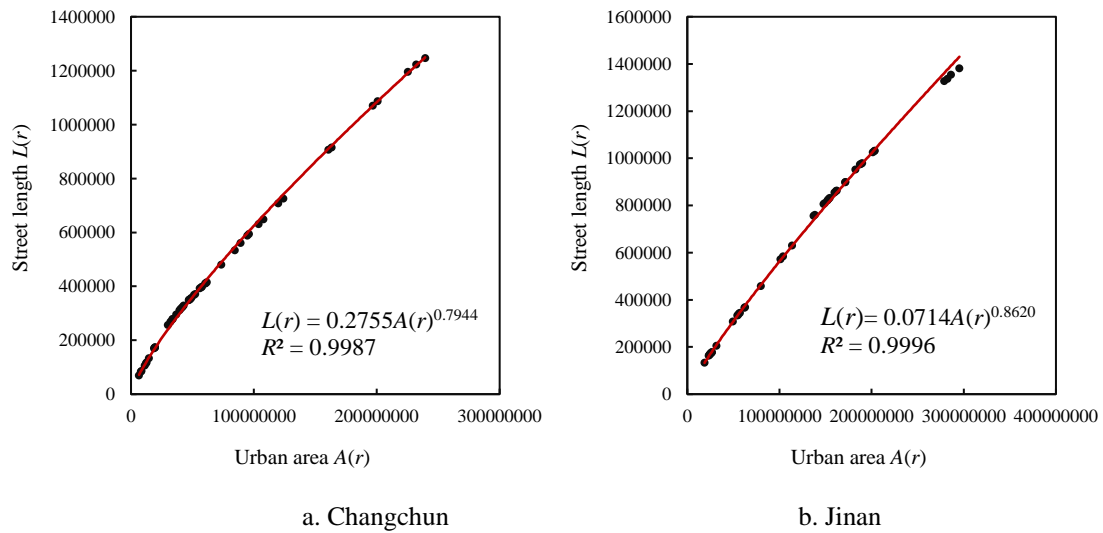

a. Changchun                 b. Jinan

**Figure 1 The allometric scaling relations between urban area and total street length of two Chinese cities defined in real space (2011)**

[**Note**: Different searching radius $r$ yields different urban boundaries for a city. A set of urban areas $A(r)$ and corresponding street lengths $L(r)$ can be extracted in terms of different urban envelopes.]

## 3.2 Case of P-space

The second example is the longitudinal allometric growth relation between urban population size and built-up area based on time series data. This type of allometry reflects the phase space (P-space) of urban geographical systems. Two Chinese cities, Beijing and Shanghai, are taken as cases (Chen and Feng, 2017). For a city, urban population and urban built-up area data can be extracted every year. In principle, the relationships between urban population, $P_t$, and urban area, $A_t$, follow the law of allometric growth. The scaling exponent is the ratio of the fractal dimension of urban form to the fractal dimension of urban population in geographical phase space. Unfortunately, owing to the unstable statistical caliber of Chinese cities, the allometric scaling relationships based on time series exhibit significant variability. Nevertheless, this example can be utilized to show the allometry growth defined in phase space (Figure 2).

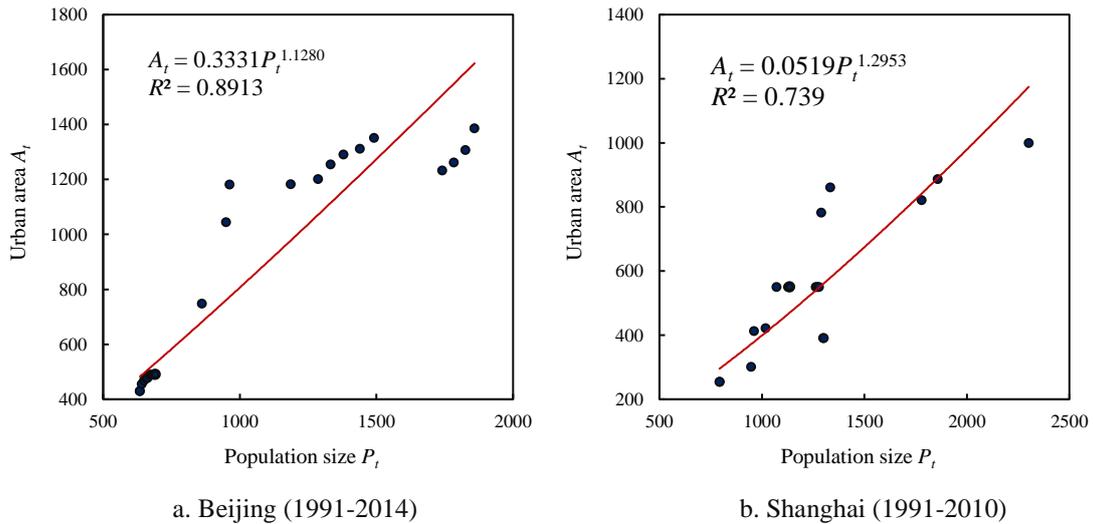

a. Beijing (1991-2014)  b. Shanghai (1991-2010)

**Figure 2 The allometric scaling relationships between population size and area of built district of Chinese cities defined in phase space**

[**Note:** Only comparable data from 1991 to 2014 are available. Beijing lacks data on urban area for 2010. Shanghai lacks urban area data for 2008 and 2009. From 2010 to 2014, the urban area data of Shanghai did not change.]

## 3.3 Case of O-space

The third example is the transversal allometric relation between urban population size and built-up area based on cross-sectional data. This type of allometry is associated with rank-size distribution and reflects the order space (O-space) of urban systems. Chinese cities in two years, 2011 and 2014, are taken as cases (Chen and Feng, 2017). Suppose that the rank of urban population size is $k$, the city size of the $k$th city is $P_k$, and the corresponding urban area is $A_k$. The relationships between urban population and urban area follow the rank-size allometric scaling law (Figure 3). The scaling exponent is the ratio of the fractal dimension of urban form to the fractal dimension of urban population in geographical order space.

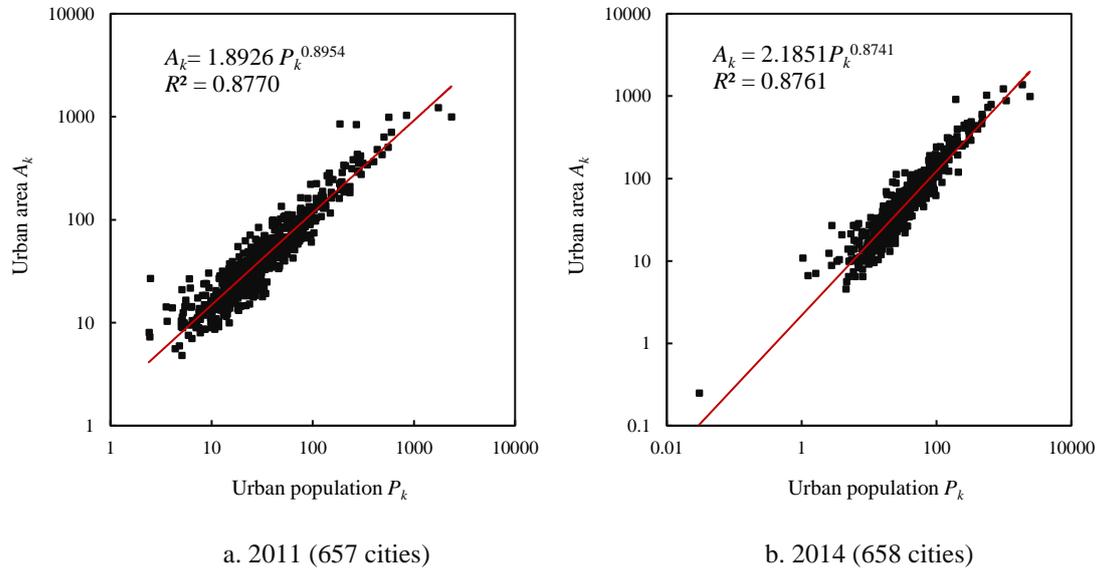

a. 2011 (657 cities)                    b. 2014 (658 cities)

**Figure 3 The rank-size allometric scaling relationships between urban population and urbanized area of Chinese cities of two years defined in order space**

[**Note:** Only comparable data of Chinese 657 to 658 cities from 1991 to 2014 are available.]

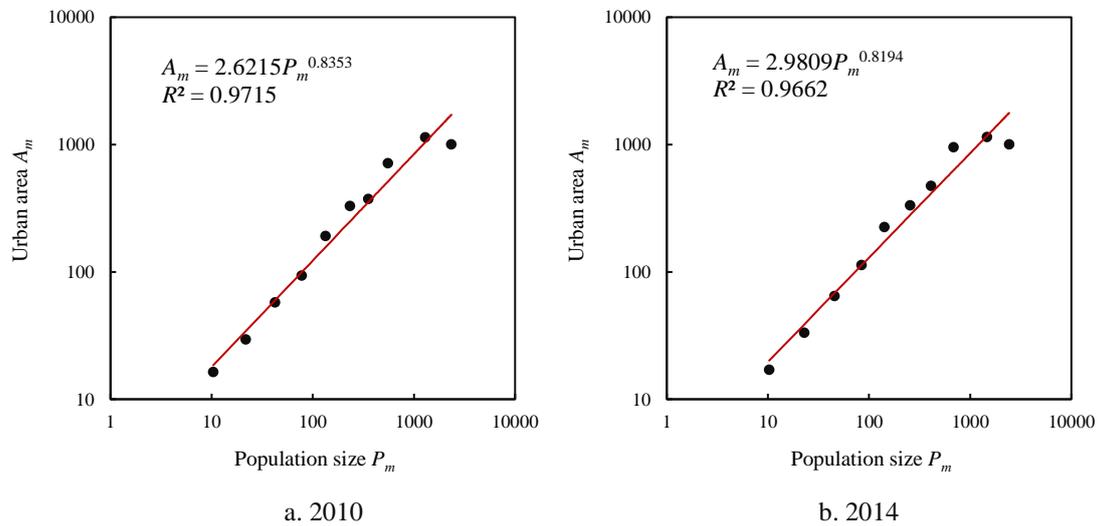

a. 2010                    b. 2014

**Figure 4 The hierarchical allometric scaling relationships between population size and area of built district of Chinese cities of two years defined in order space (10 levels)**

[**Note:** The hierarchical allometric scaling relation is equivalent to the rank-size allometric scaling relation because Zipf's law reflects a self-similar hierarchy with cascade structure.]

The cross-sectional allometry can be equivalently transformed into hierarchical allometry of cities. The hierarchical allometry also reflects the order space of urban geographical systems. Suppose that the level order of urban hierarchy is numbered as $m$, the average city size of the $m$th level is $P_m$, and the corresponding average urban area is $A_m$. The relationships between average urban population and urban area follow the hierarchical allometric scaling law (Figure 4). The scaling exponent is also the ratio of two fractal dimension of geographical order space.

# 4. Questions and discussion

Dimension is a measurement in space indicative of length, width, or height. Generally speaking, dimension implies the magnitude of something in a particular direction, e.g., length or width or height. In mathematics and science, dimension is a spatial concept, which is used to describe points, lines, and solids. In Euclidean geometry, the dimensions of points, lines, areas, and cubes are 0, 1, 2, 3, respectively. In analytic geometry, a dimension represents one of three Cartesian coordinates that determine a position in space. If we talk about the dimensions of an object or place, we always refer to its aspects, sizes, or proportions. In Euclidean geometry, dimension is apriori and cannot provide much spatial information for scientific research due to the dimension value of each Euclidean object is known. However, in fractal geometry, things are different. Fractal dimension is empirical and no longer a known quantity, but an unknown quantity to be measured. Thus, fractal dimension can provide useful information for spatial analysis. Two basic principles, which were known in Hellenic times, are important for understanding fractal dimension. One is the *principle of dimension uniqueness*. For a given object (e.g., a line), or an aspect of an object (e.g., perimeter or area of a circle), the dimension is unique. The other is the *principle of dimension consistency*. As indicated above, one measure (e.g., length, area) is in proportion to another measure (e.g., area, volume) if and only if the two measures share the same dimension.

In geospatial analysis, we have encountered and still face a series of problems to be solved. Power laws can be found in geographical world everywhere. However, the power exponent is hard to be explained by means of traditional mathematical notions in many cases (Table 4). The typical difficult problems are the relationship between the allometric exponent of urban area *vs* population growth and the distance exponent of gravity models. Suppose that the dimension of urban area is $d$=2, and the dimension of urban population is $d$=3. According to the principle of dimension consistency, we have a geometric measure relation between urban area and population such as $A=aP^{2/3}$. However, a large number of observational datasets do not lend support to this measure relation based on Euclidean dimension. Generally speaking, the allometric scaling exponent values come between 2/3 and 1, and approach 0.85 (Chen, 2010; Chen and Xu, 1999; Louf and Barthelemy, 2014). In other words, urban form and growth are fractal patterns and process. If so, it is easy to explain the values of the allometric scaling exponent. Suppose that a fractal city is defined in a 2-dimensional space. Assuming the dimension of urban form is $D=1.7$, and the dimension of urban population is $d=2$, we have $b=D/d=0.85$.

If we employ an inverse power law as an impedance function of spatial interaction, the gravity model will encounter a difficult problem of dimension. It is hard to interpret the experimental results of distance exponent value using Euclidean geometry. According to the principle of dimension consistency, the distance exponent is supposed to be an integer or an integer ratio. However, the calculation results based on the observation data are arbitrary values varying from 0 to 4. In fact, distance exponent $\sigma$ proved to the fractal dimension of city size, $D_p$, or the product of Zipf exponent, $q$, and the fractal dimension of central place network, $D_f$. That is, we have $\sigma=D_p=qD_f$. Today, many such problems can be solved by using the ideas from fractals. However, the phenomenon of fractal dimension values violating the principle of dimension uniqueness is impossible to be interpreted yet. For the same city at a given time, we can obtain a fractal dimension value from spatial measurement, and we can get another fractal dimension value from time series. The only solution to the problem of dimension uniqueness is to distinguish one type of geographical space from another type of geographical space. Different types of geographical space correspond to different types of fractal dimension, which depend on different types of observational data and calculation methods (Table 4). On the other hand, different types of fractal dimension can be employed to make different types of spatial analysis for geographical systems.

**Table 4 Several typical scientific conundrums associated with dimensions in geography**

| Principle | Meaning | Parameter | Traditional explanation | New explanation |
|---|---|---|---|---|
| **Dimension consistency** | Measure X is proportional to measure Y if and only if X and Y | Scaling exponent of allometry | A ratio of one Euclidean dimension to another Euclidean dimension | A ratio of one fractal dimension to another fractal dimension |
| | share the same dimension value | Distance exponent of gravity model | An integer or fraction based on integers | Fractal dimension or fractal dimension ratio |
| **Dimension uniqueness** | The dimension of measure X is unique | Scaling exponent based on spatial data | Integer dimension or ratio of integer dimensions | Fractional dimension or ratio of fractional dimensions |
| | | Scaling exponent based on time series data | Integer dimension or ratio of integer dimensions | Fractional dimension or ratio of fractional dimensions |
| | | Scaling exponent based on cross-sectional data | Integer dimension or ratio of integer dimensions | Fractional dimension or ratio of fractional dimensions |

# 5. Conclusions

So far, we have had four paradigms for scientific research, that is, mathematical theory, laboratory experiment, computer simulation, and data-intensive computing. The four paradigms represents mathematical method, controlled experimental method, simulation method, and computational method, respectively. Among these four paradigms, mathematical method is the basic and very important paradigm. Anyway, scientific research comprises two correlated processes: description and understanding. No exact description, no correct understanding. Mathematical modeling is the precondition for effective description and deep understanding. There are three difficult problems against mathematical modeling in scientific research: spatial dimension, time lag (response delay), interaction (coupling). The three parts method of geographical space is helpful to the effective mathematical modeling of geographical phenomena. The introduction of fractal dimension is helpful for spatial characterization of geographical systems. Only when the concept of geospatial space is clarified can fractal dimension be effectively used. Based on three types of allometric scaling of cities, geographical space were divided into three categories: real-space (R-space), phase-space (P-space), and order space (O-space). The real space can be described with box dimension and radial dimension, the phase space can be depicted by correlation dimension, and the order space can be characterized by similarity dimension. In the future, new geo-spatial theory and analytical methodologies can be developed on the basis of three-space concepts.

## Acknowledgements

This research was sponsored by the National Natural Science Foundation of China (Grant No.

42171192). The support is gratefully acknowledged.

## Appendix--Allometric growth of urban and rural population in India

The law of allometric growth was originally introduced into social science to describe the proportional relationship between urban and rural population in several countries. But urban and rural allometry is an approximate relationship for urbanization. If a country's urbanization level is low, the allometric scaling relation can be used to approximately characterize urban-rural population proportion. When the level of urbanization of the country reaches a certain critical value, this allometric relationship will break down. From 1901 to 2011, the level of urbanization in India was not too high (Table A). Indian census data shows that the relationship between urban and rural population in India approximately follow the allometry growth law (Figure A). This allometric process reflects the geographical behavior of India's urbanization in the phase space (P-space).

**Table A Census data of urban and rural population and urbanization level in Indian (1901-2011)**

| Year | Rural population | Urban population | Total population | Level of urbanization (%) | Urban-rural ratio |
|------|------------------|------------------|------------------|---------------------------|-------------------|
| 1901 | 212,544,454 | 25,851,873 | 238,396,327 | 10.8441 | 0.1216 |
| 1911 | 226,151,757 | 25,941,633 | 252,093,390 | 10.2905 | 0.1147 |
| 1921 | 223,235,046 | 28,086,167 | 251,321,213 | 11.1754 | 0.1258 |
| 1931 | 245,521,249 | 33,455,989 | 278,977,238 | 11.9924 | 0.1363 |
| 1941 | 274,507,283 | 44,153,297 | 318,660,580 | 13.8559 | 0.1608 |
| 1951 | 298,644,156 | 62,443,934 | 361,088,090 | 17.2933 | 0.2091 |
| 1961 | 360,298,168 | 78,936,603 | 439,234,771 | 17.9714 | 0.2191 |
| 1971 | 439,045,675 | 109,113,977 | 548,159,652 | 19.9055 | 0.2485 |
| 1981 | 523,866,550 | 159,462,547 | 683,329,097 | 23.3361 | 0.3044 |
| 1991 | 628,836,076 | 217,551,812 | 846,387,888 | 25.7036 | 0.3460 |
| **2001** | 741,660,293 | 285,354,954 | 1,027,015,247 | 27.7849 | 0.3848 |
| **2011** | 833,087,662 | 377,105,760 | 1,210,193,422 | 31.1608 | 0.4527 |

**Source**: http://www.censusindia.net/results/eci14_page2.html; http://en.wikipedia.org/wiki/Census_of_India,_2011.

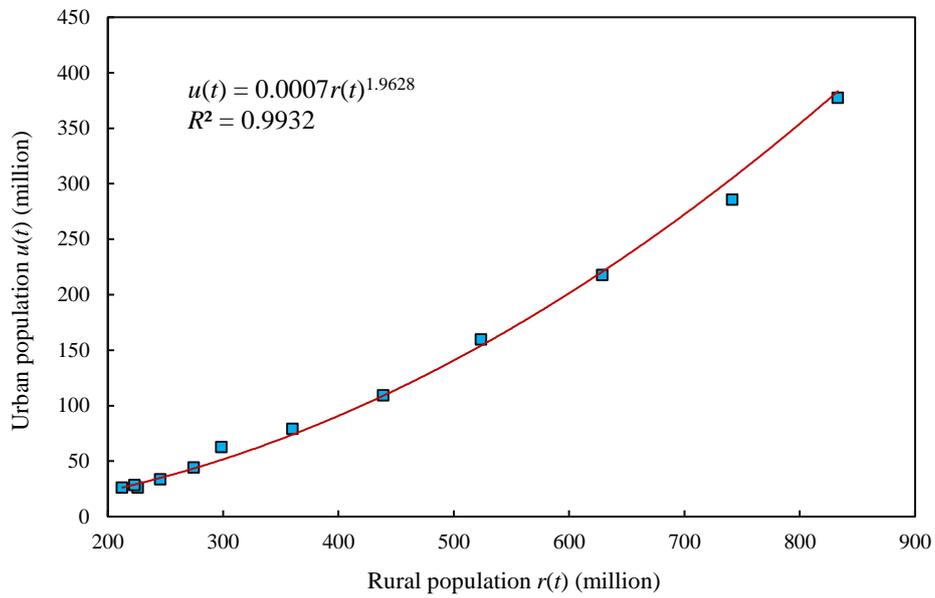

**Figure A The allometric relation between rural and urban population in India (1901-2011)**